\documentclass{article}%
\usepackage{amssymb}
\usepackage{amsmath}
\usepackage{amsfonts}
\usepackage{graphicx}%
\providecommand{\U}[1]{\protect\rule{.1in}{.1in}}
\providecommand{\U}[1]{\protect\rule{.1in}{.1in}}

\begin{document}

\title{The muon g-2 in a $SU(7)$ left-right symmetric model with mirror fermions}
\author{H. Ch\'{a}vez and J. A. Martins Sim\~{o}es\thanks{e-mail: simoes@if.ufrj.br}\\Instituto de F\'{\i}sica \\Universidade Federal do Rio de Janeiro, Brazil}
\maketitle

\begin{abstract}
We have studied a left-right symmetric model with mirror fermions based in a
grand unified SU(7) model in order to account for the muon anomaly. The Higgs sector 
of the model contains two Higgs doublets  and the hierarchy condition $\upsilon_{L}\ll\upsilon_{R}$ 
can be achieved by using two additional Higgs singlets, one even and other odd under
 $\mathcal{D}$-parity. We show that there is a wide range of values for the
mass parameters of the model that is consistent with the  $g-2$ lepton anomalies.
 Radiative correction to the mass of the ordinary fermions are shown to be small.

\end{abstract}

\section{Introduction}

Some years ago it was recognized that the measurements of $a_{l}=\frac
{g_{l}-2}{2}$ for leptons (commonly referred to as the lepton anomaly) can be
an interesting  window to discover New Physic beyond the Standard Model.

The recent theoretical results for the electron anomaly are now known at order
$\alpha^{5}$ \cite{Kinoshita} in its QED contribution and at two-loops for the
electroweak corrections \cite{Czarnecki} for the muon anomaly. The present
values are: $\Delta a_{e}\simeq(1.24\pm0.95)\times10^{-11},$ for the electron
and $\Delta a_{\mu}\simeq(2\pm2)\times10^{-9},$ for the muon. In fact, the
results reported by the Muon $(g-2)$ Collaboration \cite{Roberts} combined
with the most recent theoretical calculations have shown that there remains a
discrepancy with the SM theoretical calculations at a confidence level from
$0.7\sigma$ to $3.2$ $\sigma$, according to the values chosen for the hadronic
contributions. Using $e+e-$ data, the SM prediction for the muon $g_{\mu}-2$
deviates from the present experimental value \cite{Yndurain}\cite{Nyffeler}%
\cite{Melnikov}\cite{Davier}\cite{Passera} by $2\sigma-3\sigma$ , if the
hadronic light-by-light contribution is used instead of the hadronic $\tau$
decay data . Among all contributions yielding corrections to the muon anomaly,
the hadronic contributions are less accurate due to the hadronic vacuum
polarization effects in the diagrams which use data inputs coming from the
$e^{+}e^{-}$ annihilation cross section and the hadronic $\tau-$ decay. It is
not clear at present whether the value from $\tau-$ decay data can be improved
much further due to the difficulty in evaluating more precisely the effect of
isospin breaking.

In fact, these measurements have provided the highest accuracy for the
predictions of theories for strong, weak and electromagnetic interactions
because they have reached a fabulous relative precision of 0.5 parts per
million (ppm) in the determination of $a_{\mu}$. However, if this discrepancy
for the muon anomaly remains, it is possible that we are under a window open
for New Physics at a high energy scale, $\Lambda.$ The study of the muon
anomaly becomes relevant because it is more sensitive to interactions that are
not predicted in the SM but that can be reached at the CERN large hadron
collider (LHC) with $\sqrt{s}=14\,TeV$ .

On the theoretical side, if we take into account the effects of virtual
massive particles in the diagrams contributing to the lepton anomaly, the
corrections to the anomalies are expected to be of the order $\Big({\frac{m_{\mu}}{m_{e}}%
}\big)^{2}\sim4\times10^{4}$ for the muon, and of the order $\Big({\frac
{m_{\tau}}{m_{e}}}\Big)^{2}\sim1.2\times10^{7}$ for the tau. The same huge
enhancement factor would also affect the contributions coming from degrees of
freedom beyond the SM, so that the measurement of the $\tau-$ anomaly would
represent the best opportunity to detect new physics. Unfortunately, the very
short lifetime of the $\tau$- lepton which, precisely because of its high
mass, can also decay into hadronic states, makes such a measurement impossible
at present. This is the reason for the emphasis on the muon anomaly.

Many models beyond the Standard Model have been proposed in order to explain
the discrepancy of the muon $g_{\mu}-2$. This is the case for example in
$E_{6}$ GUT models \cite{Rizzo}\cite{Helder} , supersymmetry \cite{susy} and
left - right (L-R) models with mirror fermions \cite{Czikor}. For a review of
models for New Physic see Ref.\cite{marciano}.

In this paper we have studied a left-right symmetric model with mirror
fermions in the $SU(7)$ context using a minimal set of Higgs fields that
consists in two doublets and two singlets. One essential ingredient in our
model is the incorporation of $\mathcal{D} -$parity to induce the breaking of
$SU(2)_{R}$ through the vacuum expectation value (v.e.v.) of an odd Higgs
relative to $\mathcal{D}-$parity \cite{R.N.Mohapatra}. We expect that the
breaking scale of $SU(2)_{R}$ will be not very far from the electroweak scale, let us
say $\sim$ $TeV$ due to introduction of an even $\mathcal{D}-$parity Higgs
singlet which gain a v.e.v. in the GUT scale before the $\mathcal{D}-$parity
odd Higgs.

Our paper is organized as follow. In Section 2 we introduce our left-right
symmetrical model with mirror fermions based in a $SU(7)$ GUT. In Section 3 we
study the Higgs sector and its interactions with charged leptons that is
relevant for the lepton anomaly and its radiative mass. In Section 4 we
analyze the constraints on the relevant parameters of the model. In Section 5
we present our conclusions.

\section{A left-right symmetric model with mirror fermions in SU(7)}

Left-right symmetric models are expected to be a natural consequence of
$SU(3)_{C}\otimes SU(2)_{L}\otimes SU(2)_{R}\otimes U(1)_{Y}.$ This extension
of the standard model is also expected to be a sub-group of some grand unified
model. The fermionic content of the fundamental representations can vary, but
an economical choice consists of mirror fermions, related by a parity
symmetry. Mirror fermions are suggested for example in $SO(2n)$, $SO(2n+1)$
\cite{mirror2}, $SU(n)$ \cite{mirror3} $(n>5)$ and $E_{8}$ \cite{mirror4}
unifications models, as well as direct product group such as $SU(5)\otimes
SU(5)$ \cite{mirror5}.

In the L-R model with mirror fermions the particle content is described in
Table 1 for the two first families of fermions.
\begin{gather*}
\text{\textbf{Table 1}}\\%
\begin{tabular}
[c]{|c|c|}\hline
\textbf{Ordinary fermions} & \textbf{Mirror fermions }\\\hline
$%
\begin{array}
[c]{c}%
l_{L}=\dbinom{\nu_{e}}{e}_{L},\dbinom{\nu_{\mu}}{\mu}_{L}\sim(\mathbf{1,2,1,}%
-1\mathbf{\,)}\\
\mathbf{\ }e_{R},\mu_{R}\sim(\mathbf{1,1,1},-2)\\
\mathbf{\ }\nu_{eR},\nu_{\mu R}\sim(\mathbf{1,1,1},0)\\
\dbinom{u}{d}_{L},\dbinom{c}{s}_{L}\sim(\mathbf{3,2,1,}1/3\mathbf{\,)}\\
u_{R},c_{R}\sim(\mathbf{3,1,1,}4/3\mathbf{\,)}\\
d_{R},s_{R}\sim(\mathbf{3,1,1,}-2/3\mathbf{\,)}%
\end{array}
$ & $%
\begin{array}
[c]{c}%
L_{R}=\dbinom{N_{E}}{E}_{R},\dbinom{N_{M}}{M}_{R}\sim(\mathbf{1,1,2,}%
-1\mathbf{\,)}\\
\mathbf{\ }E\,_{L},\mathbf{\ }M_{L}\sim(\mathbf{1,1,1},-2)\\
N_{EL},N_{ML}\sim(\mathbf{1,1,1},0)\\
\dbinom{U}{D}_{R},\dbinom{C}{S}_{R}\sim(\mathbf{3,1,2,}1/3\mathbf{\,)}\\
U_{L},C_{L}\sim(\mathbf{3,1,1,}4/3\mathbf{\,)}\\
D_{L},S_{L}\sim(\mathbf{3,1,1,}-2/3\mathbf{\,)}%
\end{array}
$\\\hline
\end{tabular}
\\
\text{Content of \ the first two families of ordinary fermions with its mirror
partners }\\
\text{and quantum\ numbers under }SU(3)_{C}\otimes SU(2)_{L}\otimes
SU(2)_{R}\otimes U(1)_{Y}.
\end{gather*}

In order to justify our choice of $SU(7)$ as the unification group, some points should be observed. 
First, as $SU(3)_{C}\otimes SU(2)_{L}\otimes
U(1)_{Y}$ is a maximal sub-group of $SU(5),$ then $SU(3)_{C}\otimes
SU(2)_{L}\otimes SU(2)_{R}\otimes U(1)_{Y}$ $\subset SU(5)\otimes
SU(2)_{R}\subset SU(7)$. In fact \cite{Slansky} $SU(5)\otimes SU(2)\otimes
U(1)_{X}$ \ is a maximal sub-group of \ $SU(7)$ and we can assume $SU(2)$ to
have the right chirality $SU(2)_{R}$.

A second point is that the mass terms of leptons $\overline{l_{eL}}\chi
_{L}e_{R}$ require Higgs representations $\chi_{L}\sim(\mathbf{1,2,1,}%
1\mathbf{\,).}$ Similarly the mass terms of the mirror partners $\overline
{L_{ER}}\chi_{R}E_{L},$ require $\chi_{R}\sim(\mathbf{1,1,2,}1\mathbf{\,).} $
Mixing terms of the type $\overline{e_{R}}S_{D}E_{L},$ $\overline{\nu_{R}%
}S_{D}N_{EL}$ need $S_{D}\sim$ $(\mathbf{1,1,1,}0\mathbf{\,).}$ Mass terms of
the Majorana type $\overline{l_{eL}}\widetilde{\chi_{L}}N_{EL}^{C}$ need
$\widetilde{\chi_{L}}\sim$ $(\mathbf{1,2,1,}-1\mathbf{\,)}$ and $\overline
{L_{ER}}\widetilde{\chi_{R}}\nu_{eR}^{C}$ \ need $\widetilde{\chi_{R}}\sim$
$(\mathbf{1,1,2,}-1\mathbf{\,)}$ in order to give mass to neutrinos. The
$\overline{N_{EL}^{C}}S_{M}N_{EL}$ and $\overline{\nu_{eR}^{C}}S_{M}\nu_{eR} $
terms are possible with $S_{M}\sim(\mathbf{1,1,1},0).$ Now, let us search for
the representations of $\chi_{L,R},$ $S_{D}$ and $S_{M}$ in the $SU(7)$
context \cite{SU(7)}. The multiplet fermions \cite{K. Yamamoto} are into the
anomaly free combination\footnote{We are using $\{\}$ for the $SU(7)$
components.} $\mathbf{\{1\}\oplus\{7\}\oplus\{21}^{\ast}\mathbf{\}\oplus
\{35\}}$ corresponding to the spinor representation $\mathbf{64}$\textbf{\ }of
$SO(14)$ into \ which $SU(7) $ is embedded. In the previous multiplets,
$\left\{  \mathbf{21}\right\}  $ is a 2-fold, $\left\{  \mathbf{35}\right\}  $
is a 4-fold and $\left\{  \mathbf{7}\right\}  $ is 6-fold of totally
antisymmetric tensors. \ 

Let us note that $\mathbf{64}$ can contain two families of ordinary fermions
with its respective mirror partners, for example the electron and muon
families as it is showed in Table 1. The other families can be incorporated
into other $\mathbf{64\ }$spinorial representation. The branching rules for
each component of the spinorial representation under $SU(5)\otimes SU(2)_{R}$
, are \cite{Jihn E. Kim}: \textbf{\ }
\begin{gather}
\mathbf{\{35\}}=[\mathbf{10}^{\ast}\mathbf{,1}]\oplus\lbrack\mathbf{10,2}%
]\oplus\lbrack\mathbf{5,1}],\nonumber\\
\mathbf{\{21\}}=[\mathbf{10,1}]\oplus\lbrack\mathbf{5,2}]\oplus\lbrack
\mathbf{1,1}],\\
\mathbf{\{7\}}=[\mathbf{5,1}]\oplus\lbrack\mathbf{1,2}],\nonumber
\end{gather}
and under $SU(3)_{C}\otimes SU(2)_{L}\otimes SU(2)_{R}\otimes U(1)_{Y}$
\begin{gather}
\mathbf{\{35\}}=%
\genfrac{}{}{0pt}{0}{\underbrace{\mathbf{(1,1,1},\mathbf{-}2\mathbf{)}}%
}{e_{R}}%
\oplus%
\genfrac{}{}{0pt}{0}{\underbrace{\mathbf{(3,1,1},4/3\mathbf{)}}}{u_{R}}%
\oplus%
\genfrac{}{}{0pt}{0}{\underbrace{\mathbf{(3,2,1},1/3)}}{\dbinom{c}{s}_{L}}%
\mathbf{\oplus}%
\genfrac{}{}{0pt}{0}{\underbrace{\mathbf{(1,1,1},-2\mathbf{)}}}{E_{L}}%
\oplus\nonumber\\%
\genfrac{}{}{0pt}{0}{\underbrace{\mathbf{(1,1,1},-2\mathbf{)}}}{M_{L}}%
\oplus%
\genfrac{}{}{0pt}{0}{\underbrace{\mathbf{(3,1,1,}4/3)}}{U_{L}}%
\mathbf{\oplus}%
\genfrac{}{}{0pt}{0}{\underbrace{\mathbf{(3,1,1,}4/3)}}{C_{L}}%
\mathbf{\oplus}%
\genfrac{}{}{0pt}{0}{\underbrace{\mathbf{3,1,1},-2/3\mathbf{)}}}{s_{R}}%
\oplus\nonumber\\%
\genfrac{}{}{0pt}{0}{\underbrace{\mathbf{(1,2,1,-}1\mathbf{)}}}{\dbinom
{\nu_{e}}{e}_{L}}%
\oplus%
\genfrac{}{}{0pt}{0}{\underbrace{\mathbf{(3,1,2},1/3)}}{\dbinom{U}{D}_{R}}%
\mathbf{\oplus}%
\genfrac{}{}{0pt}{0}{\underbrace{\mathbf{\mathbf{(3,1,2},}%
1/3\mathbf{\mathbf{)}}}}{\dbinom{C}{S}_{R}}%
\mathbf{,}%
\end{gather}%
\begin{align}
\mathbf{\{21}^{\ast}\mathbf{\}}  &  =%
\genfrac{}{}{0pt}{0}{\underbrace{\mathbf{(1,1,1},-2)}}{\mu_{R}}%
\mathbf{\oplus}%
\genfrac{}{}{0pt}{0}{\underbrace{\mathbf{(3,1,1},4/3)}}{c_{R}}%
\mathbf{\oplus}%
\genfrac{}{}{0pt}{0}{\underbrace{\mathbf{(3,2,1},1/3)}}{\dbinom{u}{d}_{L}}%
\mathbf{\oplus}%
\genfrac{}{}{0pt}{0}{\underbrace{\mathbf{(1,1,2},-1)}}{\dbinom{N_{E}}{E}_{R}}%
\mathbf{\oplus}\nonumber\\
&
\genfrac{}{}{0pt}{0}{\underbrace{\mathbf{(1,1,2},-1)}}{\dbinom{N_{M}}{M}_{R}}%
\mathbf{\oplus}%
\genfrac{}{}{0pt}{0}{\underbrace{\mathbf{(3,1,1},-2/3)}}{D_{L}}%
\mathbf{\oplus}%
\genfrac{}{}{0pt}{0}{\underbrace{\mathbf{(3,1,1},-2/3\mathbf{)}}}{S_{L}}%
\oplus%
\genfrac{}{}{0pt}{0}{\underbrace{(\mathbf{1,1,1},0)}}{N_{ML}}%
,
\end{align}

\begin{align}
\mathbf{\{7\}}  &  =%
\genfrac{}{}{0pt}{0}{\underbrace{\mathbf{(1,2,1},-1)}}{\dbinom{\nu_{\mu}}{\mu
}_{L}}%
\mathbf{\oplus}%
\genfrac{}{}{0pt}{0}{\underbrace{\mathbf{(3,1,1,}-2/3)}}{d_{R}}%
\mathbf{\oplus}%
\genfrac{}{}{0pt}{0}{\underbrace{\mathbf{(1,1,1},0\mathbf{)}}}{N_{EL}}%
\mathbf{\mathbf{\oplus}}%
\genfrac{}{}{0pt}{0}{\underbrace{\mathbf{\mathbf{(1,1,1},}0\mathbf{\mathbf{)}%
}}}{N_{ML}}%
\mathbf{.}\\
\{\mathbf{1\}}  &  \mathbf{=}%
\genfrac{}{}{0pt}{0}{\underbrace{(\mathbf{1,1,1,}0)}}{\mathbf{\ }\nu_{eR}}%
\mathbf{\ }.
\end{align}
From the product $\{\mathbf{63}\}\otimes\{\mathbf{63}\}=\{\mathbf{1}%
\}_{s}\oplus\{\mathbf{63}\}_{s}\oplus\{\mathbf{63}\}_{a}\oplus...,\ $where the
index indicate symmetric (s) or antisymmetric (a), we obtain the Higgs
representations producing the mass terms for the fermions in the spinorial
multiplet $\{\mathbf{63}\}=\mathbf{\{7\}\oplus\{21}^{\ast}\mathbf{\}\oplus
\{35\}}$ of $SU(7).$ With the help of the branching rules (1) - (5), we take
\begin{gather}
\chi_{L}\sim\mathbf{\{7}^{\ast}\mathbf{\}\supset(1,2,1,}1\mathbf{),\quad}%
\chi_{R}\sim\mathbf{\{21\}\supset(1,1,2,}1\mathbf{),}\\
S_{D}\sim\{\mathbf{21\}\supset\,}(\mathbf{1,1,1,}0),\quad S_{M}\sim
\{\mathbf{1\}\sim(1,1,1,}0\mathbf{).}%
\end{gather}
Finally we can have the following breaking chain with two singlets and two
Higgs doublets :%

\begin{gather}
SU(7)\ \underrightarrow{S_{M}}\ SU(5)\otimes SU(2)_{R}\otimes\mathcal{D}%
\ \underrightarrow{S_{D}}\ G_{SM}\otimes SU(2)_{R}\nonumber\\
\ \underrightarrow{\chi_{R}}\ SU(3)_{C}\otimes SU(2)_{L}\otimes U(1)_{Y}%
\ \underrightarrow{\chi_{L}}\ SU(3)_{C}\otimes U(1)_{e.m}\ .
\end{gather}

A fundamental problem in left-right models is to satisfy the condition
$v_{R}>>v_{L}$. This can be done by introducing the concept of $\mathcal{D}%
$-parity \cite{R.N.Mohapatra,HS} The component of $\phi^{\alpha\beta
}=\{\mathbf{21\}}$ $,$ $(\alpha,\beta=1-7)$ that breaks $\mathcal{D}$-parity
is given by $S_{D}=\phi^{67}$ which is odd under $\mathcal{D}$-parity
\cite{Kim} and $S_{M}$ is a $SU(7)$ singlet that conserves $\mathcal{D}$-parity.

We can write an $SU(7)$ invariant Higgs potential which incorporates the
$\mathcal{D}$-parity effect as
\begin{gather}
\mathcal{L}=\mu^{2}\{\mathbf{7\}}\times\{\mathbf{7}^{\ast}\}+\lambda_{\chi
}(\{\mathbf{7\}}\times\{\mathbf{7}^{\ast}\})^{2}+m_{D}^{2}\{\mathbf{21\}}%
^{2}+\nonumber\\
\eta_{D}\{\mathbf{21\}}^{3}+\lambda_{D}\{\mathbf{21\}}^{4}+m_{M}%
^{2}\{\mathbf{1\}}^{2}+\eta_{M}\{\mathbf{1\}}^{3}+\lambda_{M}\{\mathbf{1\}}%
^{4}+M_{D}\{\mathbf{21\}}(\{\mathbf{7\}}\times\{\mathbf{7}^{\ast
}\})+\nonumber\\
M_{M}\{\mathbf{1\}}(\{\mathbf{7\}}\times\{\mathbf{7}^{\ast}\})+\lambda
(\{\mathbf{1\}}\times\{\mathbf{21\}})(\{\mathbf{7\}}\times\{\mathbf{7}^{\ast
}\})+\nonumber\\
(\varepsilon_{D}\{\mathbf{21\}}^{2}+\varepsilon_{M}\{\mathbf{1\}}%
^{2})(\{\mathbf{7\}}\times\{\mathbf{7}^{\ast}\})+\varkappa\lbrack
\{\mathbf{7\}}^{4}+\{\mathbf{7}^{\ast}\}^{4}].
\end{gather}
Let us note that the term $\lambda(\{\mathbf{1\}}\times\{\mathbf{21\}}%
)(\{\mathbf{7\}}\times\{\mathbf{7}^{\ast}\})$ is possible if the interactions
of $\{\mathbf{1\}}\times\{\mathbf{21\}}$ and $\{\mathbf{7\}}\times
\{\mathbf{7}^{\ast}\}$ are mediated by a gauge boson in the $\{\mathbf{21\}}$
representation of $SU(7).$

\section{Couplings in the Higgs Sector and g-2}

\subsection{The Higgs potential of a L-R model}

There are two ways of breaking parity spontaneously: the first is to identify
the discrete symmetry $Z_{2}$ that interchanges the groups $SU(2)_{L}$ \ and
$SU(2)_{R}$ of the Lorentz group $O(3,1)$ as the parity operator $\mathcal{P}
$ , which allows the parity symmetry of the Higgs bosons to be $\chi
_{L}\ \ \underleftrightarrow{\mathcal{P}}\ \chi_{R}$ and also $W_{L}%
\ \ \underleftrightarrow{\mathcal{P} }\ W_{R}$. Thus, when $SU(2)_{R}$ is
broken in the symmetric L-R model, parity is also broken. The second
possibility for a spontaneously breaking of the parity symmetry is through the
v.e.v. of an odd scalar field that conserves the L-R symmetry. In this case it
is not possible to have $\chi_{L}\ \ \underleftrightarrow{\mathcal{P}}%
\ \chi_{R}$ if in the model there are complex Yukawa couplings. This type of
parity is called $\mathcal{D}$-parity which is a generator of groups that
contain the product $SU(2)_{L}\otimes SU(2)_{R}$ as a subgroup. This second
possibility is very important because allow $\left\langle \chi_{L}%
\right\rangle \ll\left\langle \chi_{R}\right\rangle $ with different coupling
constants for $SU(2)_{L}$ and $SU(2)_{R}$ and different masses for these Higgs fields.

Our model for the scalar potential includes two Higgs doublets and two Higgs
singlets. These singlets and doublets transforms under $\mathcal{D}$-parity as
$S_{M}\ \underleftrightarrow{\mathcal{D}}\ S_{M}$ , $S_{D}%
\ \underleftrightarrow{\mathcal{D}}-S_{D}$ and $\chi_{L}\ \underleftrightarrow
{\mathcal{D}}\ \chi_{R}$ , if in the model there is no CP violation or complex
Yukawa couplings. In this case, $\mathcal{P}$ and $\mathcal{D}$-parity can be
indistinctly considered . Let us suppose the following invariant potential
under $G_{3221}=SU(3)_{C}\otimes SU(2)_{L}\otimes SU(2)_{R}\otimes U(1)_{B-L}$
for the Higgs fields%

\begin{gather}
V(\chi_{L},\chi_{R},S_{D},S_{M})=\mu^{2}({\chi_{L}^{\dagger}}\chi_{L}%
+{\chi_{R}^{\dagger}}\chi_{R})-\lambda_{\chi}({\chi_{L}^{\dagger}}\chi
_{L}+{\chi_{R}^{\dagger}}\chi_{R})^{2}-m_{D}^{2}S_{D}^{2}-\nonumber\\
\eta_{D}S_{D}^{3}-\lambda_{D}S_{D}^{4}-m_{M}^{2}S_{M}^{2}-\eta_{M}S_{M}%
^{3}-\lambda_{M}S_{M}^{4}+M_{D}S_{D}({\chi_{R}^{\dagger}}\chi_{R}-{\chi
_{L}^{\dagger}}\chi_{L})+\nonumber\\
M_{M}S_{M}({\chi_{L}^{\dagger}}\chi_{L}+{\chi_{R}^{\dagger}}\chi_{R})+\lambda
S_{D}S_{M}({\chi_{R}^{\dagger}}\chi_{R}-{\chi_{L}^{\dagger}}\chi
_{L})+\nonumber\\
(\varepsilon_{D}S_{D}^{2}+\varepsilon_{M}S_{M}^{2})({\chi_{L}^{\dagger}}%
\chi_{L}+{\chi_{R}^{\dagger}}\chi_{R})-\varkappa(({\chi_{L}^{4})}^{\dagger}%
{+}\chi_{L}^{4}+({\chi_{R}^{4})}^{\dagger}{+}\chi_{R}^{4}).
\end{gather}
Our motivation to write this potential is the fact that $S_{M}$ and $S_{D}$
are not necessarily into the same irreducible multiplet of Higgs fields. In
consequence it is also possible a mixing between these fields. If this is the
case, when $\left\langle S_{M}\right\rangle =s_{M}$\ and $\left\langle
S_{D}\right\rangle =s_{D}$ the potential responsible for the Higgs masses for
the fields $\chi_{L}$ and $\chi_{R}$\ is%

\begin{align}
V_{\text{\textbf{mass}}}(\chi_{L},\chi_{R})  &  =(\mu^{2}+\varepsilon_{D}%
s_{D}^{2}+\varepsilon_{M}s_{M}^{2}+M_{M}s_{M})(\left\vert \chi_{L}\right\vert
^{2}+\left\vert \chi_{R}\right\vert ^{2})+\nonumber\\
&  (M_{D}s_{D}+\lambda s_{D}s_{M})(\left\vert \chi_{R}\right\vert
^{2}-\left\vert \chi_{L}\right\vert ^{2}),
\end{align}
from which we find the mass terms,
\begin{align}
m_{R}^{2}  &  =\mu^{2}+\varepsilon_{D}s_{D}^{2}+\varepsilon_{M}s_{M}^{2}%
+M_{M}s_{M}+M_{D}s_{D}+\lambda s_{D}s_{M},\\
m_{L}^{2}  &  =\mu^{2}+\varepsilon_{D}s_{D}^{2}+\varepsilon_{M}s_{M}^{2}%
+M_{M}s_{M}-M_{D}s_{D}-\lambda s_{D}s_{M}.
\end{align}
Now we impose the hierarchy condition in the previous equations such that
$m_{R}^{2}$ $\ll s_{D}^{2}\ll s_{M}^{2}$. In this limit we can now have
$\left\langle \chi_{L}\right\rangle =\upsilon_{L}\sim m_{L}\sim100GeV$ and,
let us say; $\ \left\langle \chi_{R}\right\rangle =\upsilon_{R}\sim m_{R}%
\sim10TeV\gg\upsilon_{L}.$ It is necessary to indicate that $\upsilon_{L}$
breaks the electroweak symmetry and $\upsilon_{R}$ breaks the L-R symmetry
close to the TeV scale. It also must be noted that if $S_{D}$\ and $S_{M}$ are
into the same multiplet, the mixing terms in the potential possibility will
be absent.

Let us now suppose that there is no CP violation and that all v.e.v.
are considered to be real: $\left\langle \chi_{L}\right\rangle =\dbinom
{0}{\upsilon_{L}},$ $\left\langle \chi_{R}\right\rangle =\dbinom{0}
{\upsilon_{R}}.$ Then it is possible to show that the minimum conditions for
the potential are given by
\begin{gather}
\frac{\partial V}{\partial\upsilon_{L}}=2\upsilon_{L}[\mu^{2}-2\lambda_{\chi
}(\upsilon_{L}^{2}+\upsilon_{R}^{2})-M_{D}s_{D}+M_{M}s_{M}-\lambda s_{D}
s_{M}+\nonumber\\
\varepsilon_{D}s_{D}^{2}+\varepsilon_{M}s_{M}^{2}-4\varkappa\upsilon_{L}
^{2}]=0,\ \\
\frac{\partial V}{\partial\upsilon_{R}}=2\upsilon_{R}[\mu^{2}-2\lambda_{\chi
}(\upsilon_{L}^{2}+\upsilon_{R}^{2})+M_{D}s_{D}+M_{M}s_{M}+\lambda s_{D}
s_{M}+\nonumber\\
\varepsilon_{D}s_{D}^{2}+\varepsilon_{M}s_{M}^{2}-4\varkappa\upsilon_{R}
^{2}]=0,
\end{gather}
From these equations we have
\begin{equation}
\upsilon_{L}\frac{\partial V}{\partial\upsilon_{R}}-\upsilon_{R}\frac{\partial
V}{\partial\upsilon_{L}}=4\upsilon_{L}\upsilon_{R}[M_{D}s_{D}+\lambda
s_{D}s_{M}-2\varkappa(\upsilon_{R}^{2}-\upsilon_{L}^{2})]=0
\end{equation}
As we require non trivial solutions so that $\upsilon_{L}\neq$ $\upsilon
_{R}\neq0,$ we obtain the desired hierarchy
\begin{equation}
\upsilon_{R}^{2}-\upsilon_{L}^{2}=\frac{s_{D}(M_{D}+\lambda s_{M})}%
{2\varkappa}.
\end{equation}
An important point to be noted in the previous equation is that the effect of
the breaking due to the singlet $S_{M}$ is sub-dominant with relation to
$S_{D}$ which breaks $\mathcal{D}$-parity when $\left\langle S_{D}%
\right\rangle =s_{D}$ . Additionally, if $\ s_{D}=0$ the $\mathcal{D}$-parity
is conserved and also the L-R symmetry\ producing $\upsilon_{R}=\upsilon_{L}$
as expected. Thus, we have shown that in our potential there is a possibility
to construct models producing an hierarchy between the breaking scale of
$SU(2)_{R}$ and the electroweak scale simply by using two Higgs singlets to
generate the minimum of the potential. The crucial point in this sense is the
inclusion of the mixing term $\lambda S_{D}S_{M}({\chi_{R}^{\dagger}\chi}%
_{R}{-}\chi_{L}^{\dagger}\chi_{L})$ which is possible if $\ S_{M}$ and $S_{D}$
are into different irreducible representations. In the same way as in the
previous term, also the term $M_{D}S_{D}({\chi_{R}^{\dagger}}\chi_{R}%
-{\chi_{L}^{\dagger}}\chi_{L})$ breaks the L-R symmetry . It is fundamental
also to fine tune the parameters of the model at the radiative level to assure
that $\upsilon_{R}$ do not destabilizes the $\upsilon_{L}$ value. Thus, from
equations (12) - (15) we must have
\begin{equation}
m_{L}^{2}-2(\lambda_{\chi}+2\varkappa)\upsilon_{L}^{2}=2\lambda_{\chi}%
\upsilon_{R}^{2}\,,
\end{equation}

\subsection{The lepton couplings and $g-2 $}

\bigskip An interesting model with mirror fermions based in the gauge group
$SU(2)_{L}\otimes SU(2)_{R}\otimes U(1)_{Y}$ with a minimal set of Higgs
fields was elaborated in \cite{NOS}. The Lagrangian is constructed using two
Higgs doublets $\chi_{L},\chi_{R}$ that satisfy the parity transformation
$\chi_{L}\quad\underleftrightarrow{P}\quad\chi_{R}$ , and two Higgs singlets,
the first of which is coupled to Dirac terms - $S_{D}$ - and the other that
couples to Majorana terms - $S_{M}$. This is a general approach to activate
the see-saw mechanism for neutrino masses. In other L-R models \cite{SIR-1}%
\cite{BRA} with only two doublets the minimum for the vacuum appear at
$\upsilon_{L}=0,$ which is phenomenologically useless. It has also been shown
by using a variational method \cite{SIR-2} that this vacuum is unstable and
that $\upsilon_{L}$ could gain a small value, in comparison to $\upsilon_{R},$
when Higgs fields are coupled to fermion fields. The other possible solution
is the inclusion of bi-doublets, increasing the number of fundamental
parameters in the scalar sector. In the present approach, with two Higgs
doublets and two Higgs singlets, the vacuum is stable as shown
recently\cite{HS}.

The Lagrangian containing terms relevant for the anomalous magnetic moment of
the electron ( or muon ) is given by
\begin{equation}
\mathcal{L}_{\text{\textbf{Fch}}}\mathcal{=\,}f(\overline{l_{L}}\chi_{L}%
e_{R}+\overline{l_{R}}\chi_{R}E_{L})+f^{\,\prime}\,\overline{e_{R}}E_{L}%
S_{D}+h.c.
\end{equation}
The Higgs sector that breaks the symmetry is
\begin{gather}
\chi_{L}=\frac{1}{\sqrt{2}}\left(
\begin{array}
[c]{c}%
0\\
\upsilon_{L}+X_{L}%
\end{array}
\right)  ,\qquad\chi_{R}=\frac{1}{\sqrt{2}}\left(
\begin{array}
[c]{c}%
0\\
\upsilon_{R}+X_{R}%
\end{array}
\right)  ,\nonumber\\
S_{D}=\frac{1}{\sqrt{2}}\left(  s_{D}+X_{D}\right)  ,
\end{gather}
where $\upsilon_{L},$ $\upsilon_{R}$ and $s_{D}$ are the vacuum expectation
values and $X_{L},X_{R}$\ and $X_{D}$ are the respective\ neutral Higgs
fields. The fermion mass terms are
\begin{equation}
\left(
\begin{array}
[c]{cc}%
\overline{e_{L}} & \overline{E_{L}}%
\end{array}
\right)  \frac{1}{\sqrt{2}}\left(
\begin{array}
[c]{cc}%
f\upsilon_{L} & 0\\
f\,^{\prime}s_{D} & f\upsilon_{R}%
\end{array}
\right)  \left(
\begin{array}
[c]{c}%
e_{R}\\
E_{R}%
\end{array}
\right)  +h.c.
\end{equation}
\ A rotation between the fermion fields will diagonalize the mass matrix
\begin{gather}
e_{L}=c_{L}e_{L}^{0}+s_{L}E_{L}^{0},\quad e_{R}=c_{R}e_{R}^{0}-s_{R}E_{R}%
^{0}\,,\nonumber\\
E_{L}=-s_{L}e_{L}^{0}+c_{L}E_{L}^{0},\quad E_{R}=s_{R}e_{R}^{0}+c_{R}E_{R}%
^{0}\,,
\end{gather}
where \ $s_{L,R}=\sin\theta_{L,R}\,,$ $c_{L,R}=\cos\theta_{L,R}$. The weak
eigenstates are $e_{L,R}$ and $E_{L,R}$ while $e_{L,R}^{0}$ and $E_{L,R}^{0}$
are the mass eigenstates. Then, the terms of the previous Lagrangian
contributing to the magnetic moment and CP conserving are
\begin{gather}
\mathcal{L}_{\text{\textbf{eE}}}\mathcal{=}\frac{f}{\sqrt{2}}[(s_{L}%
c_{R}\overline{E_{L}^{0}}e_{R}^{0}-c_{L}s_{R}\overline{e_{L}^{0}}E_{R}%
^{0})X_{L}+(s_{R}c_{L}\overline{e_{R}^{0}}E_{L}^{0}-c_{R}s_{L}\overline
{E_{R}^{0}}e_{L}^{0})X_{R}]\nonumber\\
+\frac{f\,^{\prime}}{\sqrt{2}}(c_{L}c_{R}\overline{e_{R}^{0}}E_{L}^{0}%
+s_{L}s_{R}\overline{E_{R}^{0}}e_{L}^{0})X_{D}+h.c.
\end{gather}

The Feynman diagram producing a new contribution to the anomalous magnetic
moment of leptons consistent with the requirement of a well defined Higgs
potential is shown in Fig.1.

\begin{center}%
\begin{center}
\includegraphics[
height=1.6725in,
width=2.4007in
]%
{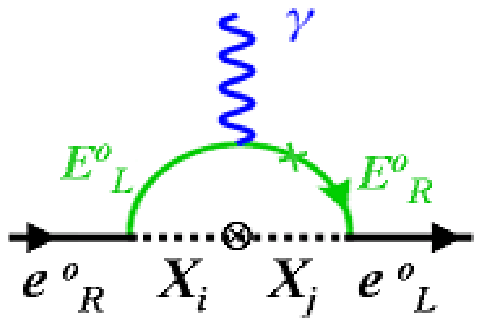}%
\\
Fig. 1 Generic Feynman diagram contributing to the lepton anomaly. The Higgs
fields are in the mass eigenstates basis. The index is $i\neq j=L,R,D.$%
\end{center}

\end{center}

Note the mixing term between Higgs fields that arise from the potential (10).
This mixing is crucial to give a finite radiative mass to leptons (the same
diagram as in Fig.1, but without the photon line). Figure 1 is the relevant
contribution for the anomalous magnetic moment in the hierarchy $\upsilon
_{D}\gg\upsilon_{R}\gg\upsilon_{L}$ with $M_{D}\gg M_{R}\gg M_{L}$. This
contribution for the electron case is, with the additional condition $m_{E}\gg
m_{e}$ for the mirror partner of the electron $E$, is given by \cite{Leveille}%
\cite{Kephart}%

\begin{equation}
\Delta a_{e}=\frac{\xi_{e}}{16\pi^{2}}\frac{m_{e}}{m_{E}}\times\frac{\left(
1-z_{e}^{2}\right)  ^{2}-2z_{e}^{2}\left(  1-z_{e}^{2}\right)  -2z_{e}^{4}%
\ln(z_{e}^{2})}{2\left(  1-z_{e}^{2}\right)  ^{3}},
\end{equation}
where $\ z_{e}=\frac{M_{L}}{m_{E}}$ and the parameter $\xi_{e}$ is function of
$f,\theta_{L,R}$ and of the mixing angle between $X_{L}$ and $X_{R}$ that can
be easily obtained from the Lagrangian $\mathcal{L}_{\text{\textbf{eE}}}$. The
corresponding contribution to the electron radiative mass is given by
\cite{Kephart}\cite{marciano}%
\begin{equation}
m_{e}^{\text{\textbf{1-loop}}}\simeq\frac{\xi_{e}}{16\pi^{2}}m_{E}\left[
\frac{M_{L}^{2}}{m_{E}^{2}-M_{L}^{2}}\ln\left(  \frac{m_{E}^{2}}{M_{L}^{2}%
}\right)  -\frac{M_{R}^{2}}{m_{E}^{2}-M_{R}^{2}}\ln\left(  \frac{m_{E}^{2}%
}{M_{R}^{2}}\right)  \right]  \,.
\end{equation}
We will use the very small contribution to $m_{e}^{\text{\textbf{1-loop}}}$
and to the anomalous magnetic moment to obtain constrains over $\xi_{e}%
,M_{L},M_{R}$ and $m_{E}.$ We obtain analogous expressions for the muon.

\section{Bounds from $g-2$ and radiative masses.}

In this section we use two simple arguments in order to obtain bounds over the
parameters in our model. The first one is  the value of the leptons anomaly and the second is the small value for the radiative mass. Let us take $\ -10^{-11}\lesssim\Delta a_{e}%
\lesssim3\times10^{-11}$ for the electron anomaly. In Fig. 2 we show the
possible range of values for $\xi_{e}$ as a function of $M_{L}/m_{E}$ for two
different cases: $m_{E}=100GeV$ and $200GeV$. In this paper we consider the mass parameter that fixes the Standard Model Higgs boson to be in the range $47GeV\leqslant M_{L}\leqslant200GeV$\ \cite{Erler}.

We have also obtained constraints for $\xi_{e}$ from the radiative mass given in equation(25)
as is shown in Fig.3. A small radiative mass for the electron is possible in
our model. For example, by taking $m_{e}^{\text{\textbf{1-loop}}}=0.05MeV$,
$m_{E}=100GeV$ \ with $47GeV\leqslant M_{L}\leqslant200GeV$ \ we found
$-2\times10^{-5}\lesssim\xi_{e}\lesssim-4.5\times10^{-5}$ in the range of
values $6\leqslant\frac{M_{R}}{m_{E}}\lesssim10$. This is showed in  Fig.3a.
For the case  $m_{E}=200GeV$ \ with the same electron radiative mass we have the results of
Fig.3b. In both cases, the range of values of $\xi_{e}$ is compatible with the
values coming from the electron anomaly, as is showed in Fig. 2a and  Fig.2b. Thus, a Higgs SM with mass in the
indicated range and its mirror partner with a mass $600GeV\leqslant M_{R}\leqslant1TeV$ can 
account for the electron anomaly and give a small electron radiative mass. 

\begin{center}%
\begin{center}
\includegraphics[
trim=0.000000in 0.000000in 0.056972in 0.054934in,
height=4.5163cm,
width=11.6289cm
]%
{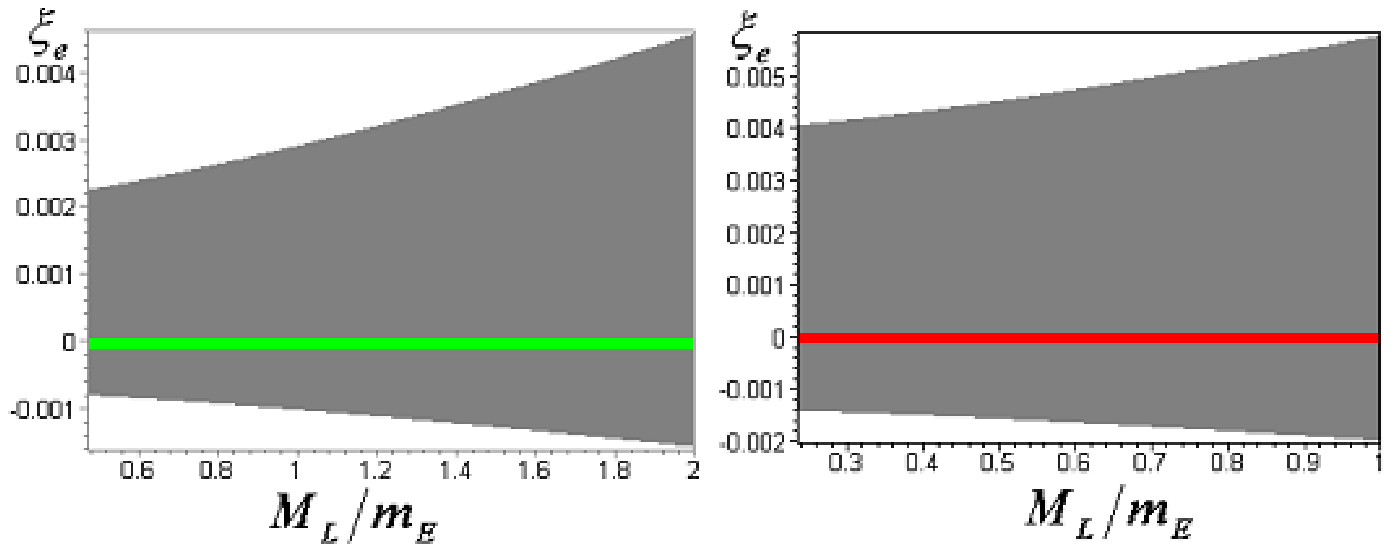}%
\\
Fig. 2 Allowed values for  $\xi_{e}$ coming from the electron $g_{e}-2$ as a
function of $M_{L}/m_{E}$ for two cases: a) $m_{E}=100\,GeV$ and b)
$m_{E}=200\,GeV.$%
\end{center}

\begin{center}
\includegraphics[
height=4.3471cm,
width=11.6597cm
]%
{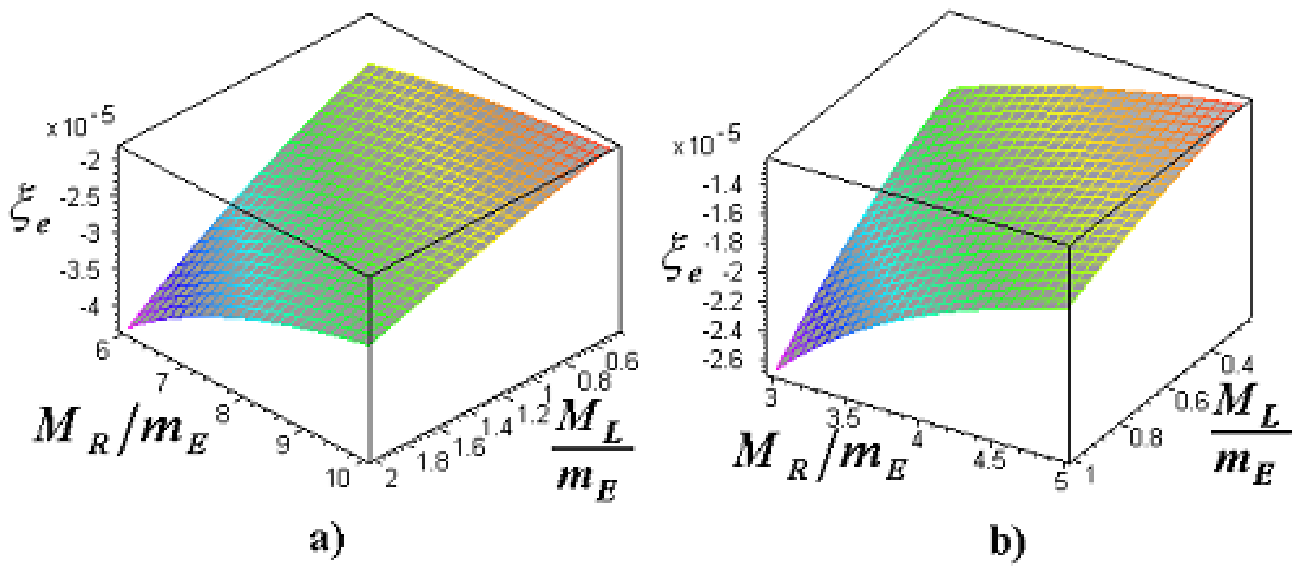}%
\\
Fig. 3 Range of allowed values for $\xi_{e}$ from the radiative mass contribution, taking $m_{e}^{\text{\textbf{1-loop}}}=0.05MeV$, as a function of
$M_{R}/m_{E}$ and $M_{L}/m_{E}$ for: a) $m_{E}=100GeV$ and b) \ $m_{E}%
=200GeV$.
\end{center}

\end{center}

\bigskip 

For the muon case let us assume the value $-2\times10^{-9}\lesssim\Delta a_{\mu
}\lesssim6\times10^{-9}$ for the anomalous magnetic moment. The
range of values for  $\xi_{\mu}$ coming from muon anomaly is shown in Fig.4. For
the radiative muon mass we have the results shown in Fig. 5.

\begin{center}
\begin{center}
\includegraphics[
height=4.6744cm,
width=11.7102cm
]%
{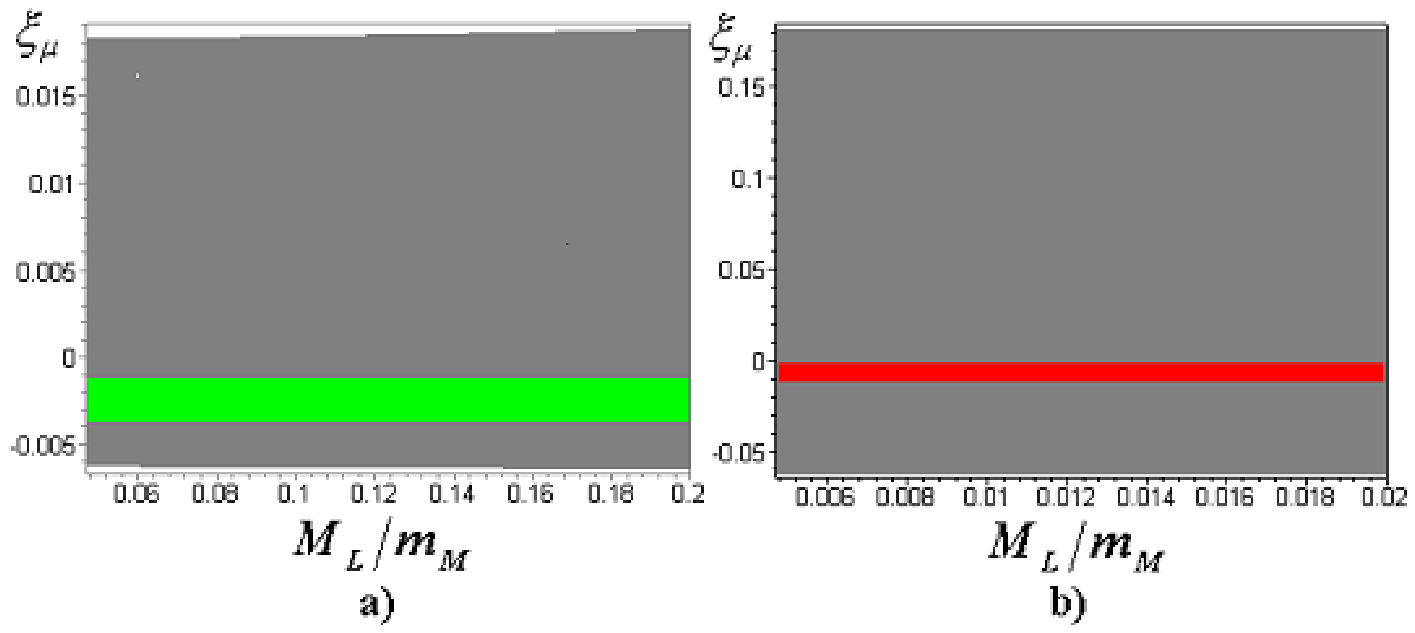}%
\\
Fig. 4 Range of values for  $\xi_{\mu}$ coming  from  $g_{\mu}-2$ as a function of
$M_{L}/m_{M}$, for: a) $m_{M}=1TeV$ and b) $m_{M}=10TeV.$%
\end{center}

\end{center}

The shadowed area in  Fig. 4a gives a range of values $-1.5\times
10^{-3}\leqslant\xi_{\mu}\leqslant-3.5\times10^{-3}$ coming from the muon radiative mass
 for $M_{R}=600GeV-1TeV$ , $47GeV\leqslant M_{L}%
\leqslant200GeV$ and a small radiative mass $m_{\mu}^{\text{\textbf{1-loop}}%
}=10MeV$. This  is compatible with the values for  $\ \xi_{\mu}$ coming from of
muon anomaly.

\begin{center}%
\begin{center}
\includegraphics[
height=4.8304cm,
width=11.607cm
]%
{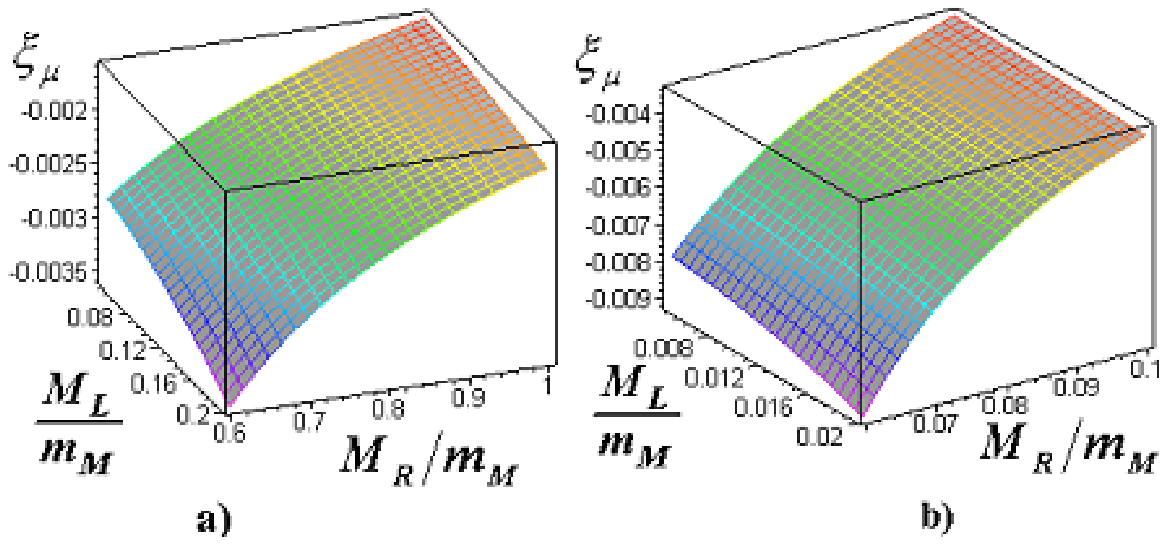}%
\\
Fig. 5 Range of values of $\xi_{\mu}$ from the muon radiative mass, as a function of $M_{R}/m_{M}$ and $M_{L}/m_{M}$. We consider  $m_{\mu
}^{\text{\textbf{1-loop}}}=10MeV$ \ for: a) $m_{M}=1TeV$ and b)\ $m_{M}%
=10TeV.$%
\end{center}

\end{center}

Let us notice  from  Fig.5 that it is possible to obtain smaller values for the muon radiative mass
 $1\leqslant m_{\mu}^{\text{\textbf{1-loop}}}\leqslant10MeV$
\ by taking $m_{M}=1TeV-10TeV$ \ for a Standard Model Higgs with mass between $47-200GeV$ and $M_{R}$
between $600GeV-1TeV$. These cases are totally compatible with the present value of the muon anomaly.

\section{Comments on new corrections}
Left-right symmetric models with mirror fermions will have other contributions to the muon anomaly.
\par
There are  corrections from vacuum polarization loops with $E$ and $M-$mirror, similar to the 
QED contribution. The insertion of $E-$mirror vacuum
polarization loop into the muon vertex correction, give $\Delta a_{\text{%
\textbf{v.p.l}}}$ $\simeq $ $[\frac{1}{45}\left( \frac{m_{\mu }}{m_{E}}%
\right) ^{2}+$ $\frac{1}{70}\frac{m_{\mu }^{4}}{m_{E}^{4}}\times \ln \frac{%
m_{E}}{m_{\mu }}]\left( \frac{\alpha }{\pi }\right) ^{2}$. For the range of parameters of the new mirror sector considered in this paper we have the value $ \sim  10^{-14}-10^{-13}.$  
\par
 New gauge bosons will also contribute to the muon anomaly with diagrams analogous to the standard model gauge bosons. As the new gauge bosons must have masses higher than the standard gauge bosons we will have again very small contributions.
\par
The new hadronic contribution from mirror matter will suffer the same limitations as in the standard model. One can not calculate  the hadronic contribution directly from QCD due to the non-perturbative region of parameters. The standard procedure is to use dispersion relations and experimental data from  $e^{+}e^{-}$ collisions or the pion spectral functions.  Hadronic vacuum polarization loops with new mirror quarks could 
contribute to the muon anomaly. But we would need to know the $e^{+}e^{-}$  annihilation  into mirror quark-antiquark and their consequent hadronization. Similar possibilities could occur for the hadronic $\tau -$ decay. We expect
that this new hadronic contributions from  mirror quarks will be very small in virtue
of the hierarchy masses present in our model, in comparison with the  ordinary quarks masses.

\section{Conclusions}

In conclusion, we found that in a $SU(7)$ grand unified symmetric left-right model it is possible to have 
 a Standard Model  Higgs boson  with mass between $47-200GeV$ , a new 
mirror Higgs boson with mass between $600GeV-1TeV;$ a mirror electron with
$m_{E} \simeq 100GeV$ and a mirror muon with mass $m_{M} \simeq 1TeV$. These values are compatible with
the electron and muon anomaly and give a small radiative mass contribution. These possible new high mass 
states can be searched in the LHC energies \cite{Simoes}and new high energy lepton colliders\cite{Martins}

As was showed in \cite{Helder}, the important linear relation between masses
of ordinary and mirror fermions for the electron or muon anomaly is related to
the breaking of the Weinberg symmetry \cite{Weinberg}, if the radiative
correction to the masses of the ordinary fermions are small, as it  is the  case in the present model.

\begin{description}
\item[\textbf{Acknowledgments}] We thanks CNPq and FAPERJ for financial support.
\end{description}


\begin{thebibliography}{99}                                                                                               %


\bibitem {Kinoshita}T. Kinoshita and M. Nio. Phys. Rev. D \textbf{73}%
(2006)013003, T. Aoyama, M. Hayakawa, T. Kinoshita and M. Nio, Nucl. Phys. B
\textbf{740}(2006)138.

\bibitem {Czarnecki}T. Gribook and A. Czarnecki, Phys. Rev. D \textbf{72}(2005)053016

\bibitem {Roberts}B. Lee Roberts (for the Muon (g-2) Collaborations),
hep-ex/0501012; G. W. Bennet et al., (BNL Muon g-2 Collaboration),
Phys.Rev.Lett.92 (2004) 161802; E. Sichtermann et al., (BNL Muon (g-2)
Collaboration), hep-ex/0309008, G. W. Bennet et al., Phys. Rev. Lett. 89
(2002) 129903.

\bibitem {Yndurain}J. F. de Troconiz anf F. J. Yndurain, Phys.Rev. D
\textbf{71} (2005) 073008, M. Knecht, Lect. Notes Phys. \textbf{629}(2004)37,
V. Barger, Ch. Kao, P. Langacker and Hye-Sung Lee, Phys.Lett. B \textbf{614}
(2005) 67, A. Czarnecki, Nuc.Phys. B (Proc. Suppl.) \textbf{144} (2005) 201.

\bibitem {Nyffeler}A. Nyffeler, Acta Phys. Polon. B\textbf{\ 34 }(2003) 5197.

\bibitem {Melnikov}K. Melnikov and A. Vainshtein, Phys. Rev. D 70(2004)113006.

\bibitem {Davier}M. Davier, S. Eidelman, A. H\"{o}cker and Z. Zhang, Euro.
Phys. J. C. \textbf{31} (2003) 503, T. Teubner, Eur Phys. J. C \textbf{33}
(2004) 653, A. H\"{o}cker, hep-ph/0410081, K. Hagiwara, A.D. Martin, Daisuke
Nomura and T. Teubner, arXiv:hep-ph/0312250 v3

\bibitem {Passera}M. Passera, hep-ph/0509372 v2, , M. Passera, Jour. Phys. G.:
Nucl. Part. Phys. \textbf{31} (2005) R75.

\bibitem {Rizzo}Thomas Rizzo, Phys. Rev. D \textbf{33}(1986)3329, \ I.
Vendramin, IL Nuovo Cimento, \textbf{100}(1988)757

\bibitem {Helder}Helder Chavez, Cristine N. Ferreira and Jos\'{e} A.
Helayel-Neto, Phys. Rev. D \textbf{74}(2006)033006-1, hep-ph/0410373 v2.

\bibitem {susy}J. L. Lopez, D. V. Nanopoulos and X. Wang, Phys. Rev. D
\textbf{49}(1994)366, J. Ellis, J. Hagellin and D. V. Nanopoulos, Phys. Lett.
B \textbf{116}(1982)283, G.-C. Cho, K. Hagiwara and M. Hayakawa, Phys. Lett. B
\textbf{478}(2000)231, T. Ibrahim and P. Nath, Phys. Rev. D \textbf{62}%
(2000)015004, U. Chattopadhyay, D. K. Ghosh and S. Roy, Phys. Rev. D
\textbf{62}(2000)115001, Tai-Fu Feng, Xue-Qian Li, Lin Lin, Jukka Maalampi,
He-Shan Song, hep-ph/0604171.

\bibitem {Czikor}J. Maalampi, J. T. Peltoniemi and M. Ross, Phys. Lett. B
\textbf{220}(1989)441, F. Czikor and Z. Fodor, Phys. Lett. B \textbf{287}(1992)358.

\bibitem {marciano}Andrzej Czarnecki and William Marciano, Phys. Rev. D
\textbf{64}(2001)013014.

\bibitem {R.N.Mohapatra}D. Chang, R. N. Mohapatra and M. K. Parida, Phys. Rev.
Lett. \textbf{52}(1984)1072, D. Chang, R. N. Mohapatra and M. K. Parida, Phys.
Rev. D \textbf{30}(1984)1052, D. Chang, R. N. Mohapatra, J. M. Gibson, R. E.
Marshak and M. K. Parida, Phys. Rev. D \textbf{31}(1985)1718, D. Chang and A.
Kumar, Phys. Rev. D \textbf{33}(1986)2695.

\bibitem {mirror2}K. Enqvist and J. Maalampi, Nucl. Phys. B \textbf{191}%
(1981)189, J.~Maalampi and M.~Roos, Phys.\ Rept.\ \textbf{186}(1990)53 .

\bibitem {mirror3}J. Chakrabarti, M. Popovic and R. N. Mohapatra, Phys. Rev. D
\textbf{21}(1980)3212, I. Umemura and K. Yamamoto, Phys. Lett. B \textbf{100}(1981)34

\bibitem {mirror4}I. Bars and G\"{u}naydin, Phys. Rev. Lett. \textbf{45}
(1980)859, S. M. Barr, Phys. Rev. D \textbf{37}(1988)204.

\bibitem {mirror5}M. Collie and R. Foot, Phys. Lett. B \textbf{432}(1998)134.

\bibitem {Slansky}R. Slansky, Phys. Rep. \textbf{79} (1981) 1.

\bibitem {SU(7)}N. S. Baaklini, Phys. Rev. D \textbf{21}(1980)1932, A. Umemura
and K. Yamamoto, Progress of Theor. Phys. \textbf{66}(1981)1430.

\bibitem {K. Yamamoto}K. Yamamoto, Mirror Fermions in the SU(7) GUT and Their
Effects on Flavour Changing Process, Kyoto University, unpublished, 1983.

\bibitem {Jihn E. Kim}Kyuwan Hwang and Jihn E. Kim, Phys. Lett. B \textbf{540}
(2002) 289, M. Claudson, A. Yildiz and P. H. Cox, Phys. Lett. B \textbf{97(}%
1980\textbf{)}224.

\bibitem {Kim}Jihn E. Kim, Phys. Rev. D \textbf{26} (1982)2009.

\bibitem {NOS}Y.~A.~Coutinho, J.~A.~Martins Simoes and C.~M.~Porto,
Eur.\ Phys.\ J.\ C \textbf{18}(2001)779; F.~M.~L.~Almeida, Y.~A.~Coutinho,
J.~A.~Martins Simoes, J.~Ponciano, A.~J.~Ramalho, S.~Wulck and M.~A.~B.~Vale,
Eur.\ Phys.\ J.\ C \textbf{38}(2004)115; J.~A.~Martins Simoes and J.~Ponciano,
Eur.\ Phys.\ J.\ C \textbf{32S1}(2004)91.

\bibitem {SIR-1}F.~Siringo, Phys.\ Rev.\ Lett.\ \textbf{92} (2004) 119101.

\bibitem {BRA}B.~Brahmachari, E.~Ma and U.~Sarkar,
Phys.\ Rev.\ Lett.\ \textbf{91}(2003)011801 .

\bibitem {SIR-2}F. Siringo and L. Marotta, hep-ph/0605276.

\bibitem {HS}H. Ch\'{a}vez and J. A. Martins Sim\~{o}es hep-ph/0606112 v2.

\bibitem {Leveille}Jacques P. Leveille, Nuc. Phys. B \textbf{137}(1978)63

\bibitem {Kephart}T. W. Kephart and H. P\"{a}s, Phys. Rev. D \textbf{65}(2002)093014.

\bibitem {Erler}Jens Erler, hep-ph/0607332 v2.

\bibitem{Simoes}
  Y.~A.~Coutinho, J.~A.~Martins Simoes, C.~M.~Porto and P.~P.~Queiroz Filho,
  Phys.\ Rev.\ D {\bf 57}, 6975 (1998).
  
\bibitem{Martins}
  F.~M.~L.~Almeida, Y.~A.~Coutinho, J.~A.~Martins Simoes, S.~Wulck and M.~A.~B.~do Vale,
  Eur.\ Phys.\ J.\ C {\bf 30}, 327 (2003)

\bibitem {Weinberg}S. Weinberg, The Quantum Theory of Fields (Cambridge
University Press, Cambridge, England, 1995, Vol.1, p. 520).
\end{thebibliography}
\end{document}